\documentclass{SciPost}

\hypersetup{
    colorlinks,
    linkcolor={red!50!black},
    citecolor={blue!50!black},
    urlcolor={blue!80!black}
}

\usepackage[bitstream-charter]{mathdesign}
\DeclareSymbolFont{usualmathcal}{OMS}{cmsy}{m}{n}
\DeclareSymbolFontAlphabet{\mathcal}{usualmathcal}

\fancypagestyle{SPstyle}{
\fancyhf{}
\lhead{\colorbox{scipostblue}{\bf \color{white} ~SciPost Physics Core}}
\rhead{{\bf \color{scipostdeepblue} ~Submission }}

\fancyfoot[C]{\textbf{\thepage}}
}

\usepackage{xspace}     
\usepackage{mathtools}  
\usepackage{braket}
\usepackage{nth}
\usepackage{pgfplots}

\usetikzlibrary{arrows.meta}
\usetikzlibrary{backgrounds}
\usepgfplotslibrary{patchplots}
\usepgfplotslibrary{fillbetween}
\pgfplotsset{%
    layers/standard/.define layer set={%
        background,axis background,axis grid,axis ticks,axis lines,axis tick labels,pre main,main,axis descriptions,axis foreground%
    }{
        grid style={/pgfplots/on layer=axis grid},%
        tick style={/pgfplots/on layer=axis ticks},%
        axis line style={/pgfplots/on layer=axis lines},%
        label style={/pgfplots/on layer=axis descriptions},%
        legend style={/pgfplots/on layer=axis descriptions},%
        title style={/pgfplots/on layer=axis descriptions},%
        colorbar style={/pgfplots/on layer=axis descriptions},%
        ticklabel style={/pgfplots/on layer=axis tick labels},%
        axis background@ style={/pgfplots/on layer=axis background},%
        3d box foreground style={/pgfplots/on layer=axis foreground},%
    },
    compat=1.18
}

\newcommand{\pcst}{$\textrm{pcst}^{\texttt{++}}$\xspace}
\newcommand{\ii}{\textrm{i}}
\newcommand{\identity}{\mathbb{1}}
\newcommand{\abs}[1]{\left|{#1}\right|}

\DeclareMathOperator{\im}{imag}
\newcommand{\ketbra}[2]{\ket{{#1}}\!\bra{{#2}}}
\newcommand{\kett}[1]{\ket{{#1}}\!\rangle}
\newcommand{\bbra}[1]{\langle\!\bra{{#1}}}
\newcommand{\op}{\mathcal{O}}
\newcommand{\Lindbladian}{\mathcal{L}}
\DeclareMathOperator{\sgn}{sgn}

\newcommand{\boson}{b}
\newcommand{\crealwo}{\boson_{+}^\ddagger}
\newcommand{\annilwo}{\boson_{+}^{\phantom{\ddagger}}}
\newcommand{\crearwo}{\boson_{-}^\ddagger}
\newcommand{\annirwo}{\boson_{-}^{\phantom{\ddagger}}}
\newcommand{\crealrwo}{\boson_{\pm}^\ddagger}
\newcommand{\annilrwo}{\boson_{\pm}^{\phantom{\ddagger}}}

\newcommand{\creal}[1]{\boson_{{#1}, +}^\ddagger}
\newcommand{\annil}[1]{\boson_{{#1}, +}^{\phantom{\ddagger}}}
\newcommand{\crear}[1]{\boson_{{#1}, -}^\ddagger}
\newcommand{\annir}[1]{\boson_{{#1}, -}^{\phantom{\ddagger}}}
\newcommand{\crealr}[1]{\boson_{{#1}, \pm}^\ddagger}
\newcommand{\annilr}[1]{\boson_{{#1}, \pm}^{\phantom{\ddagger}}}

\newcommand{\annirl}[1]{\boson_{{#1}, \mp}^{\phantom{\ddagger}}}
\newcommand{\crea}[1]{\boson_{{#1}}^\ddagger}
\newcommand{\anni}[1]{\boson_{{#1}}^{\phantom{\ddagger}}}

\newcommand{\kettdd}{\kett{\downarrow \downarrow}}
\newcommand{\kettud}{\kett{\uparrow \downarrow}}
\newcommand{\kettdu}{\kett{\downarrow \uparrow}}
\newcommand{\kettuu}{\kett{\uparrow \uparrow}}
\newcommand{\bbradd}{\bbra{\downarrow \downarrow}}
\newcommand{\bbraud}{\bbra{\uparrow \downarrow}}
\newcommand{\bbradu}{\bbra{\downarrow \uparrow}}
\newcommand{\bbrauu}{\bbra{\uparrow \uparrow}}

\begin{document}

\pagestyle{SPstyle}

\begin{center}{\Large \textbf{\color{scipostdeepblue}{
Effective quasiparticle conserving Lindbladians in the thermodynamic limit\\
}}}\end{center}

\begin{center}\textbf{
Lea Lenke\textsuperscript{$\star$}
and
Kai Phillip Schmidt\textsuperscript{$\dagger$}
}\end{center}

\begin{center}
Department of Physics, Staudtstra{\ss}e 7, Friedrich-Alexander-Universit\"at Erlangen-N\"urnberg, Germany
\\[\baselineskip]
$\star$ \href{mailto:email1}{\small lea.lenke@fau.de}\,,\quad
$\dagger$ \href{mailto:email2}{\small kai.phillip.schmidt@fau.de}
\end{center}

\section*{\color{scipostdeepblue}{Abstract}}
\textbf{\boldmath{
Open quantum many-body systems are commonly described by Lindblad master equations, yet the treatment of Lindbladian operators in the thermodynamic limit remains a major challenge.
We develop a framework for constructing effective quasiparticle-conserving Lindbladian operators directly in the thermodynamic limit.
Our approach extends continuous similarity transformations to non-Hermitian open quantum systems and enables the systematic block diagonalization of Lindbladians with respect to the quasiparticle number.
We formulate two complementary methods.
The first, projective continuous similarity transformations (\pcst), generalizes perturbative continuous unitary transformations to Lindblad operators so that a linked-cluster expansion allows us to obtain high-order series expansions of the infinite system.
The second, deepCST, extends directly evaluated enhanced perturbative continuous unitary transformations by combining the same quasiparticle-conserving generator with a perturbative truncation scheme that yields non-perturbative effective Lindbladian operators directly in thermodynamic limit.
As a paradigmatic example, we apply \pcst and deepCST to the dissipative transverse-field Ising chain with local dissipation.
We focus on the regime which is adiabatically connected to the limit of vanishing Ising interactions where the model is exactly solvable and exhibits a paramagnetic steady state.
We purify the Lindbladian by splitting each spin into a dimer of two sites.
A physical spin flip then corresponds to two quasiparticle excitations.
We derive and analyze the effective, quasiparticle-conserving Lindbladian in the sectors with zero to two quasiparticle excitations.
We show how the Ising interaction renormalizes the decay rates of elementary spin-flip excitations and provides the microscopic mechanism for a competition between coherent interactions and dissipation.
Our work establishes perturbative and non-perturbative continuous similarity transformations as a versatile tool for deriving effective quasiparticle pictures of open quantum many-body systems in the thermodynamic limit.
}}

\vspace{\baselineskip}

\noindent\textcolor{white!90!black}{%
\fbox{\parbox{0.975\linewidth}{%
\textcolor{white!40!black}{\begin{tabular}{lr}%
  \begin{minipage}{0.6\textwidth}%
    {\small Copyright attribution to authors.\newline
    This work is a submission to SciPost Physics.\newline
    License information to appear upon publication.\newline
    Publication information to appear upon publication.}
  \end{minipage} & \begin{minipage}{0.4\textwidth}
    {\small Received Date \newline Accepted Date \newline Published Date}%
  \end{minipage}
\end{tabular}}
}}
}


\vspace{10pt}
\noindent\rule{\textwidth}{1pt}
\tableofcontents
\noindent\rule{\textwidth}{1pt}
\vspace{10pt}


\section{Introduction}
\label{sec:intro}

Open quantum many-body systems exhibit a rich variety of non-equilibrium phenomena that have no direct counterpart in isolated systems.
Of particular interest are dissipative phase transitions, where the competition between coherent many-body dynamics and coupling to an environment gives rise to qualitatively distinct steady-state phases and critical behavior.
Besides being of fundamental interest, such transitions have become increasingly relevant experimentally, as modern quantum platforms like driven-dissipative cavity and circuit-QED systems, trapped ions, Rydberg-atom arrays, and ultracold atomic gases allow both, coherent interactions and dissipation to be engineered and controlled with high precision~\cite{CarrRitterWadeAdamsWeatherill2013,FinkDombiVukicsWallraffDomokos2017,FitzpatrickSundaresanLiKochHouck2017,LetscherThomasNiederprumFleischhauerOtt2017,RodriguezCasteelsStormeCarlonSagnesLeGratietGalopinLemaitreAmoCiutiBloch2017,WalkerFlattenHestenMintertHungerTrichetSmithNyman2018,CaiLiuJiangWuMeiZhaoHeZhangZhouDuan2022,LiClaudeBoulierGiacobinoGlorieuxBramatiCiuti2022,BenaryBaalsBernhartJiangRohrleOtt2022,BeaulieuMingantiFrascaSavonaFelicettiDiCandiaScarlino2025}.
Dissipation therefore no longer merely represents an unavoidable source of decoherence, but can constitute an essential ingredient for creating and stabilizing novel many-body states~\cite{DiehlMicheliKantianKrausBuchlerZoller2008,HarringtonMuellerMurch2022}.
Understanding dissipative phase transitions and the associated steady states and long-lived excitations is consequently an important challenge for the theory of open quantum many-body systems.

While closed quantum many-body systems are already difficult to treat theoretically because the Hilbert space grows exponentially with the system size, this scaling is even more enhanced when studying open quantum systems.
These systems are usually not exactly solvable, i.e., their eigenvalues and eigenstates cannot be calculated analytically.
This calls for approximations that are often influenced by methods developed for quantum optics, e.g. quantum trajectories~\cite{Daley2014}, or for closed quantum many-body systems, e.g. wave-function Monte Carlo methods~\cite{DalibardCastinMolmer1992,DumZollerRitsch1992,GisinPercival1992}, tensor network methods~\cite{VerstraeteGarciaRipollCirac2004,CuiCiracBanuls2015,FinazziBoiteStormeBaksicCiuti2015,WernerJaschkeSilviKlieschCalarcoEisertMontangero2016,KshetrimayumWeimerOrus2017}, variational methods~\cite{Weimer2015,OverbeckWeimer2016,OverbeckMaghrebiGorshkovWeimer2017}, mean-field theory~\cite{TomadinGiovannettiFazioGeraceCarusottoTureciImamoglu2010,DiehlTomadinMicheliFazioZoller2010}, cluster mean-field theory~\cite{JinBiellaViyuelaMazzaKeelingFazioRossini2016}, dynamical mean-field theory~\cite{PanasPasekDharQinGeisslerHafezTorbatiSorantinTitvinidzeHofstetter2019}, and numerical linked-cluster expansions~\cite{BiellaJinViyuelaCiutiFazioRossini2018}.
These methods developed from closed quantum many-body system methods are summarized in more detail in~\cite{WeimerKshetrimayumOrus2021}.
Most of them are either related to mean-field theory or they use large but finite systems.
Others like numerical linked-cluster expansions or variational techniques operate directly in the thermodynamic limit but truncating the system in spatial correlations or entanglement content, respectively.
What is largely missing, however, are quasiparticle-based approaches that systematically incorporate interactions while operating directly in the thermodynamic limit. 

In recent years, Rosso et al.~\cite{RossoIeminiSchiroMazza2020} and Schmiedinghoff et al.~\cite{SchmiedinghoffUhrig2022} successfully applied the framework of continuous unitary transformations (CUTs), also known as flow equation approach \cite{GlazekWilson1993,Wegner1994}, to Lindbladians.
Methods within this framework typically yield results on the operator level in the thermodynamic limit.
In particular, they can be formulated in terms of quasiparticles (qps) \cite{SchmiedinghoffUhrig2022}, thus taking into account proper particle interactions.
This qp picture for open quantum many-body systems described by effective Lindbladians is the general framework we continue to extend in our work.

For Hermitian many-body Hamiltonians admitting a qp description, the pc-generator \cite{KnetterUhrig2000} provides a natural choice for CUTs, as it yields a quasiparticle-conserving and thus block-diagonal effective Hamiltonian.
It is therefore suited to transform a many-body Hamiltonian into a few-body effective Hamiltonian, which can then be used to calculate the low-energy properties of the system.
If the original system is not exactly solvable, it is not possible to perform the CUT analytically.
In this case, there exist different physically motivated truncation schemes to obtain a closed set of equations.
In this work, we exclusively use perturbative truncation schemes, in order to generate perturbatively and non-perturbatively approximated effective Lindbladians.
Our methods are based on are pCUT~\cite{Mielke1998,KnetterUhrig2000} and deepCUT~\cite{KrullDrescherUhrig2012}.
The latter is, in turn, based on epCUT~\cite{KrullDrescherUhrig2012}.
All three CUTs use the pc-generator and a perturbative truncation scheme, but differ in their advantages and generate different effective Hamiltonians.
The advantage of pCUT over the other two is that it can be formulated model-independently, i.e., it does not depend on the specific form of the Hamiltonian which originates from the requirement that the unperturbed Hamiltonian must have a ladder spectrum.
The other two methods, epCUT and deepCUT, can be applied to more general systems, but they are model-dependent.
Their advantage over pCUT is that their generated effective Hamiltonians are normal-ordered, which makes reading off eigenvalues easier.
Both methods have the same perturbative truncation scheme, but while epCUT also generates a perturbative effective Hamiltonian, deepCUT generates a non-perturbative effective Hamiltonian.
This Hamiltonian can be understood as taking into account also processes of higher orders, which go beyond the perturbative order of the truncation scheme.

Because these methods transform many-body Hamiltonians into few-body Hamiltonians with low qp numbers, they are very well suited to calculate ground-state and excitation energies.
This makes them natural for calculating quantum phase diagrams because they yield results directly for the thermodynamic limit.
They have also been used for calculating bound states~\cite{WindtGruningerNunnerKnetterSchmidtUhrigKoppFreimuthAmmerahlBuchnerRevcolevschi2001,SchmiedinghoffMullerKumarUhrigFauseweh2022} and static and dynamic correlation functions~\cite{KnetterSchmidtGruningerUhrig2001,KnetterSchmidtUhrig2003b,FausewehUhrig2013}.
It is possible to use them for long-range interacting systems~\cite{FeySchmidt2016,FeyKapferSchmidt2019,AdelhardtKoziolSchellenbergerSchmidt2020,AdelhardtKoziolLangheldSchmidt2024} and to combine them with Floquet theory~\cite{VerdenyMielkeMintert2013,VoglLaurellBarrFiete2019,ThomsonMaganoSchiro2021}.
They can be made more efficient by implementing them as a graph expansion and more versatile by using white graphs~\cite{CoesterSchmidt2015}.

The treatment of Lindbladians requires different transformations than the treatment of Hamiltonians, in the context of qp-conserving CUTs mainly because they are not Hermitian.
Instead of a unitary transformation, one has to perform a continuous similarity transformation (CST).
Our first CST method, \pcst~\cite{LenkeSchellenbergerSchmidt2023}, is obtained by performing two generalizations of pCUT.
One generalization is to use the gpc-generator~\cite{SchmiedinghoffUhrig2022} instead of the pc-generator, which allows us to treat non-Hermitian operators.
The other generalization is to use multiple qp types.
These generalizations enable us to perform a high-order perturbative linked cluster series expansion, that can be efficiently implemented using a graph expansion.
Our second method, deepCST, is obtained by generalizing deepCUT to use the gpc-generator instead of the pc-generator.
To our knowledge, a deepCST expansion of a Lindbladian was not done before.
Both methods can be used in any dimension and are valid in the thermodynamic limit.

There are examples of CSTs being used to treat non-Hermitian Hamiltonians~\cite{PowalskiUhrigSchmidt2015,PowalskiSchmidtUhrig2018,LenkeMuhlhauserSchmidt2021,WaltherHeringUhrigSchmidt2023} and Lindbladians~\cite{RossoIeminiSchiroMazza2020,SchmiedinghoffUhrig2022}.
In both previous applications of CSTs to Lindbladians, the treated Lindbladians were either exactly solvable, considered on a finite cluster or rewritten as finite matrices.
Our approaches use the Lindbladians directly as they are formulated in terms of qps instead, which makes the treatment of Lindbladians more similar to the treatment of Hamiltonians.
The ground states and elementary excitations in the Hamiltonian case are those states with the lowest energies; in our treatment of Lindbladians this gets replaced by the steady states and the longest living modes.
This makes calculating the Liouvillian gap to obtain the steady state phase diagram conceptually very similar to calculating the energy gap and obtaining the quantum phase diagram of Hamiltonians.

Method development requires conceptually simple yet nontrivial benchmark models.
Analogously to the paradigmatic transverse field Ising model, there exist different open generalizations of this model in different dimensions formulated as Lindblad equations.
We are interested in the case where the dissipation is local and in the direction of the magnetic field.
Here we concentrate on the dissipative transverse-field Ising model (DTFIM) in one dimension~\cite{JoshiNissenKeeling2013,FossFeigYoungAlbertGorshkovMaghrebi2017,LenkeSchellenbergerSchmidt2023,AliKamarSeifMaghrebi2026}.
While a continuous dissipative quantum phase transition had been predicted on the mean-field level, numerical investigations have shown the absence of such a transition \cite{JoshiNissenKeeling2013}.
Let us note that this model was studied also in higher dimensions by different approaches~\cite{MaghrebiGorshkov2016,OverbeckMaghrebiGorshkovWeimer2017,JinBiellaViyuelaCiutiFazioRossini2018,JinHeIeminiFerreiraWangChesiFazio2021}.

In this work, we develop and benchmark two complementary qp-conserving approaches for interacting Lindbladians directly in the thermodynamic limit. The first one, \pcst~\cite{LenkeSchellenbergerSchmidt2023}, extends pCUT to non-Hermitian Lindblad generators and yields high-order perturbative linked-cluster expansions, while the second one, deepCST, introduced here as an extension of deepCUT, provides a non-perturbative solution of perturbatively truncated flow equations. Both approaches employ the gpc-generator~\cite{SchmiedinghoffUhrig2022} and transform the many-body Lindbladian into an effective block-diagonal form organized by qp number, thereby providing direct access to the steady state and the longest-lived decay modes in low-qp sectors. We apply both methods to the one-dimensional DTFIM and calculate the effective Lindbladian in the zero-, one-, and two-qp sectors. From the resulting qp spectra, we determine the Liouvillian gap and find it to remain finite throughout the investigated parameter regime, consistent with the absence of a dissipative phase transition in one dimension~\cite{JoshiNissenKeeling2013}. Our results establish \pcst and deepCST as complementary tools for studying steady-state phases and their elementary long-lived excitations in interacting open quantum many-body systems directly in the thermodynamic limit.

This work is organized as follows.
In Sec.~\ref{sec:model}, we introduce the Lindbladian of the dissipative transverse field Ising chain and the qps we use.
We explain \pcst in Sec.~\ref{sec:pcst} and deepCST in Sec.~\ref{sec:deepcst} and the qp-conserving structure of the effective Lindbladians in Sec.~\ref{sec:blocks}.
The results are presented in Sec.~\ref{sec:results}, where we compare the lifetimes calculated with the two methods.
Finally, we conclude in Sec.~\ref{sec:conclusion}.

\section{Dissipative transverse field Ising chain}
\label{sec:model}

\begin{figure}[bt]
  \centering
  \includegraphics[page = 1, clip, trim = 0 19.1cm 0 11.7cm, width=\textwidth]{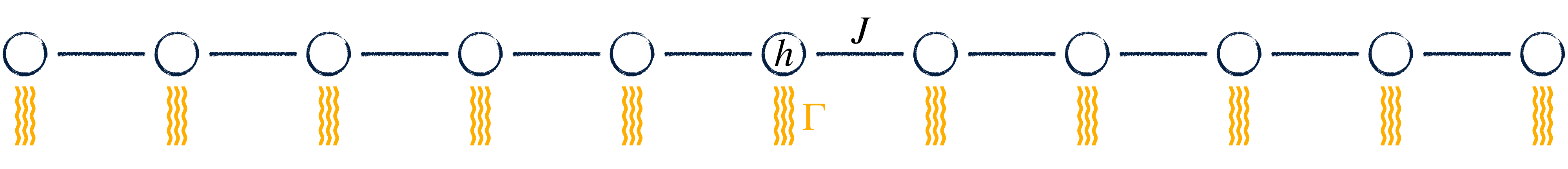}
  \caption{
    Illustration of the one-dimensional DTFIM.
    Neighboring spin-1/2 particles on a chain are coupled with an Ising interaction of strength $J$, subject to a magnetic field of strength $h$, and local dissipation with rate $\Gamma$.
  }
  \label{fig:DTFIM-chain}
\end{figure}

The Hamiltonian of the transverse field Ising model is given by
\begin{equation}
  H = h \sum_j \sigma^z_j - J \sum_{\langle i, j\rangle} \sigma^x_i \sigma^x_j,
\end{equation}
where $\sigma^\alpha_j$ are Pauli operators, $h > 0$ is the magnetic field, and $J > 0$ is the Ising interaction.
This Hamiltonian is known to have a second-order quantum phase transition at $J = h$ between a paramagnetic phase for $J < h$ and a ferromagnetic phase for $J > h$~\cite{LiebSchultzMattis1961,Pfeuty1970}.
The dissipation is introduced by the Lindblad operators
\begin{equation}
  L_j^{\phantom{-}} = \sigma^-_j,
\end{equation}
in the Lindblad equation
\begin{equation}
  \ii \dot{\rho} = [H, \rho] + \frac{\ii \Gamma}{2} \sum_j \left( 2 L_j^{\phantom{\dagger}} \rho L_j^\dagger - \{ L_j^\dagger L_j^{\phantom{\dagger}}, \rho \} \right).
\end{equation}
All terms are illustrated in Fig.~\ref{fig:DTFIM-chain}.

\begin{figure}[b]
  \centering
  \includegraphics[page = 1, clip, trim = 0 6.7cm 0 22.5cm, width=\textwidth]{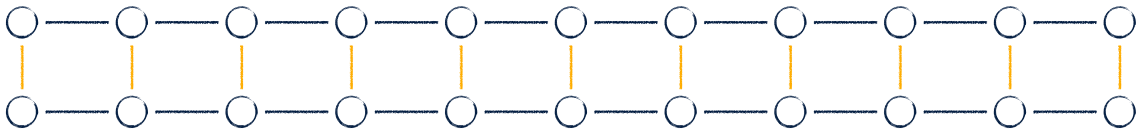}
  \caption{
    Illustration of the purified, one-dimensional DTFIM in Liouville space on an effective ladder geometry.
    Every spin-1/2 particle is represented by a rung dimer consisting of two sites, one for $\sigma_j \otimes \identity$ and one for $\identity \otimes \sigma_j$.
    It can also be seen as a representation of the Lindbladian in terms of qp, where the creation and annihilation operators $\creal{j}$ and $\annil{j}$ act on one leg and $\crear{j}$ and $\annir{j}$ on the other.
    Note that the two sites of a rung dimer $j$ are not exactly identical in both representations, since the creation operators act on both sites of a rung dimer in the original spin representation (see Eqs.~\eqref{eq:particleplus} and \eqref{eq:particleminus}).
  }
  \label{fig:DTFIM-ladder}
\end{figure}

We purify this equation and work in Liouville space, where the density operator is represented as a vector.
The transformation from operator space to Liouville space is defined by mapping its basis vectors $\ketbra{\varphi}{\psi}$ to the basis vectors $\kett{\varphi, \psi} \coloneq \ket{\varphi}\otimes\ket{\psi}^*$.
Operators are mapped to vectors via $\kett{\op} \coloneq \sum_{\ket{\varphi}, \ket{\psi}} \bra{\varphi} \op \ket{\psi} \kett{\varphi, \psi}$.
The Lindblad equation is mapped to $\kett{\ii \dot{\rho}} = \Lindbladian \kett{\rho}$, with the Lindbladian
\begin{equation}
  \begin{split}
  \Lindbladian &= h \sum_j \left( \sigma^z_j \otimes \identity - \identity \otimes \sigma^z_j \right) - J \sum_{\langle i, j\rangle} \left( \sigma^x_i \sigma^x_j \otimes \identity - \identity \otimes \sigma^x_i \sigma^x_j \right)\\
  &\hphantom{={}} + \frac{\ii \Gamma}{2} \sum_j \left( 2 \sigma^-_j \otimes \sigma^-_j - \sigma^+_j \sigma^-_j \otimes \identity - \identity \otimes \sigma^+_j \sigma^-_j \right),
  \end{split} \label{eq:Lindbladian}
\end{equation}
which is an operator acting on elements of Liouville space.
It is illustrated in Fig.~\ref{fig:DTFIM-ladder}.
Note that, because every original site is split into two sites, one physical spin flip in the original system corresponds to two nearest-neighbor spin flips on a rung dimer in Liouville space.

Solutions of the Lindbladian consist of eigenstates $\kett{\rho}$ with eigenvalue $\lambda$.
They have the time evolution $\kett{\rho(t)} = e^{- \ii \lambda t} \kett{\rho(0)}$.
The real part of $\lambda$ yields oscillatory behavior, while the non-positive imaginary part indicates the lifetime of the state.
If the imaginary part is zero the state does not decay and is called a stationary state.
The further the imaginary part is from zero, the faster the state decays.
Although the whole Lindbladian is not exactly solvable, we know that the state without any excitations is a stationary state independent of the chosen parameters.

\subsection{Unperturbed eigenbasis and quasiparticles}
\label{sec:model-unperturbed}

The magnetic field term and the dissipative term are local and commute with each other, while the Ising interaction term is non-local and does not commute with the other two terms.
This is why we work in the perturbative limit $\abs{J} \ll \abs{h}, \abs{\Gamma}$, where the Ising interaction is treated as a perturbation.
Because of the non-Hermitian nature of the dissipative term, the unperturbed parts have a mutual bi-orthonormal eigenbasis with left and right eigenstates.
In accordance with the notation of~\cite{LenkeSchellenbergerSchmidt2023}, we label them
\begin{align}
  \ket{0, 0} &\coloneq \kettdd, & \ket{+1, 1} &\coloneq \kettud, & \ket{-1, 1} &\coloneq \kettdu, & \ket{0, 2} &\coloneq \kettuu - \kettdd,\\
  \bra{0, 0}^L &\coloneq \bbrauu + \bbradd, & \bra{+1, 1}^L &\coloneq \bbraud, & \bra{-1, 1}^L &\coloneq \bbradu, & \bra{0, 2}^L &\coloneq \bbrauu.
\end{align}
The superscript $L$ indicates that the left eigenstates are not equal to the usual adjoints of the corresponding right eigenstates.
This basis can be used to define an operation similar to the usual adjoint operation, but using the left eigenstates instead.
This operation is not denoted by superscript $\dagger$, but by superscript $\ddagger$ to avoid confusion.

In order to simplify the unperturbed parts of the Lindbladian, we introduce the local hardcore-bosonic creation and annihilation operators
\begin{align}
  \crealwo &= \sigma^+ \otimes \identity - \identity \otimes \sigma^- + 2 \sigma^+ \sigma^- \otimes \sigma^-, & \annilwo &= \sigma^- \otimes \identity, \label{eq:particleplus}\\
  \crearwo &= \identity \otimes \sigma^+ - \sigma^- \otimes \identity + 2 \sigma^- \otimes \sigma^+ \sigma^-, & \annirwo &= \identity \otimes \sigma^- \label{eq:particleminus}
\end{align}
with mutual bosonic commutation relations.
Expressed in terms of the unperturbed local eigenbasis, these operators read
\begin{align}
  \crealrwo &= \ketbra{\pm 1, 1}{0, 0}^L + \ketbra{0, 2}{\mp 1, 1}^L, & \annilrwo &= \ketbra{0, 0}{\pm 1, 1}^L + \ketbra{\mp 1, 1}{0, 2}^L. \label{eq:qp-operators}
\end{align}
Here it becomes apparent why the annihilation operators $\annilrwo$ are related to the creation operators $\crealrwo$ via the adapted adjoint operation.
This is the operator language we use for the whole paper.
Note that these qps are not the ones used in~\cite{LenkeSchellenbergerSchmidt2023}, but they are related to them via
\begin{align}
  \crealrwo &= a_{\pm}^\ddagger + a_{*}^\ddagger a_{\mp}^{\phantom{\ddagger}}, & \annilrwo &= a_{\pm}^{\phantom{\ddagger}} + a_{\mp}^\ddagger a_{*}^{\phantom{\ddagger}}.
\end{align}
The unperturbed parts of the Lindbladian can be expressed in terms of these qps as
\begin{equation}
  Q = 2 h \sum_j \left( \creal{j} \annil{j} - \crear{j} \annir{j} \right) - \frac{\ii \Gamma}{2} \sum_j \left( \creal{j} \annil{j} + \crear{j} \annir{j} \right),
\end{equation}
which is equivalent to counting the two types of qps with prefactors $\pm 2 h - \ii \Gamma/2$.
In this limit, the state without any qps is the only stationary state, and the more qps are added, the shorter the lifetime of the state.
Note that, to treat a state with one physical spin flip in the original system, we need to create two qps -- one of either type.

\subsection{Perturbation}
\label{sec:model-perturbed}

The perturbation expressed in terms of these creation and annihilation operators consists of more terms than in the original formulation~\eqref{eq:Lindbladian}.
Because it makes sense for the methods we use, we group the terms according to the amount $m^\pm$ of qps that they create or annihilate.
The whole perturbation is of the form
\begin{equation}
  V = - J \sum_{(m^{+}, m^{-})} T_{(m^{+}, m^{-})}
\end{equation}
and the individual terms $T_{(m^{+}, m^{-})}$ are given in Table~\ref{tab:perturbation}.
\begin{table}[bt]
  \begin{tabular}{c|l}
    $(m^{+}, m^{-})$ & Terms in $T_{(m^{+}, m^{-})}$\\
    \hline\hline
    $(0, - 2)$ & $- 2 \annir{i} \creal{j} \annil{j} \annir{j} - 2 \creal{i} \annil{i} \annir{i} \annir{j} + 4 \creal{i} \annil{i} \annir{i} \creal{j} \annil{j} \annir{j}$\\
    \hline
    $(- 2, 0)$ & $+ 2 \annil{i} \crear{j} \annil{j} \annir{j} + 2 \crear{i} \annil{i} \annir{i} \annil{j} - 4 \crear{i} \annil{i} \annir{i} \crear{j} \annil{j} \annir{j}$\\
    \hline
    $(+ 2, 0)$ & $+ \creal{i} \creal{j}$\\
    \hline
    $(0, + 2)$ & $- \crear{i} \crear{j}$\\
    \hline
    $(0, 0)$ & $\creal{i} \annil{j} + \annil{i} \creal{j} -  \crear{i} \annir{j} - \annir{i} \crear{j}$\\
    \hline
    $(-1, -1)$ & $2 \annir{i} \crear{j} \annil{j} \annir{j} + 2 \crear{i} \annil{i} \annir{i} \annir{j} - 2 \annil{i} \creal{j} \annil{j} \annir{j} - 2 \creal{i} \annil{i} \annir{i} \annil{j}$\\
    \hline
    $(+ 1, - 1)$ & $+ \creal{i} \annir{j} + \annir{i} \creal{j} - 2 \creal{i} \creal{j} \annil{j} \annir{j} - 2 \creal{i} \annil{i} \annir{i} \creal{j}$\\
    \hline
    $(- 1, + 1)$ & $- \crear{i} \annil{j} - \annil{i} \crear{j} + 2 \crear{i} \crear{j} \annil{j} \annir{j} + 2 \crear{i} \annil{i} \annir{i} \crear{j}$
  \end{tabular}
  \caption{
    Perturbation terms of the Lindbladian of the dissipative transverse field Ising chain~\eqref{eq:Lindbladian} expressed in terms of the creation and annihilation operators~(\ref{eq:particleplus}, \ref{eq:particleminus}).
    The first column indicates the amount $m^\pm$ of in total created qps of either type when acting with the term in the second column.
  }
  \label{tab:perturbation}
\end{table}
When acting on an eigenstate of the unperturbed Lindbladian $Q$, the perturbation operator $T_{(m^+, m^-)}$ changes its eigenvalue by
\begin{equation}
  (2h - \ii \Gamma/2) m^+ + (-2h - \ii \Gamma/2) m^-.
\end{equation}

\subsection{Symmetries}
\label{sec:model-symmetries}

To make the subsequent calculations easier, we can exploit some symmetries of the Lindbladian.
Firstofall, we must use the translational symmetry
\begin{equation}
  \annilr{j} \to \annilr{j + 1},
\end{equation}
which ensures that we get a finite number of independent terms in the latter calculations.
Additionally, we take into account inversion symmetry
\begin{equation}
  \annilr{j} \to \annilr{- j}
\end{equation}
and the symmetry defined by swapping the two qp types, negating and complex conjugating the Lindbladian
\begin{equation}
  \annilr{j} \to \annirl{j}, \quad \Lindbladian \to - \Lindbladian^*.
\end{equation}
The latter two symmetries are optional, but useful, as they reduce the number of independent terms in the calculations.
This is especially helpful for the deepCST calculations, because it reduces the number of coupled differential equations we need to solve numerically.

\section{Continuous similarity transformations}
\label{sec:cut}

CSTs are generalizations of CUTs~\cite{GlazekWilson1993,Wegner1994} to non-Hermitian operators.
They are used to rotate an operator into a more suitable basis in which it has a convenient form, e.g., a block-diagonal form.
This is done by performing a similarity transformation $S(\ell)$ depending on the continuous flow parameter $\ell \in [0, \infty]$.
The transformation is defined via its infinitesimal generator $\eta(\ell)$, i.e., as the solution of
\begin{equation}
  \partial_\ell S(\ell) = - \eta(\ell) S(\ell), \label{eq:flow-S}
\end{equation}
with the initial condition $S(0) = \identity$.
Note that $\eta(\ell)$ is not necessarily anti-Hermitian, as would be the case for CUTs.
Various useful generators have been developed, depending on the desired properties of the effective operator.
Since we want to derive a qp-conserving Lindbladian and study the steady-state phase diagram, we use the gpc-generator~\cite{SchmiedinghoffUhrig2022}, which is a generalization of the pc-generator~\cite{KnetterUhrig2000} to non-Hermitian operators.
The gpc-generator is defined using matrix element notation in the eigenbasis of the unperturbed Lindbladian $Q$ as
\begin{equation}
  \eta(\ell)_{\mu \nu} = \sgn(Q_{\mu \mu} - Q_{\nu \nu}) \Lindbladian(\ell)_{\mu \nu}, \label{eq:generator}
\end{equation}
where $\Lindbladian(\ell) = S(\ell) \Lindbladian S(\ell)^{-1}$ is the flowing Lindbladian.
Because $Q$ is non-Hermitian, we use an extension of the sign function
\begin{equation}
  \sgn(z) = \begin{cases}
    0 & \textrm{for } z = 0,\\
    \frac{z^*}{\abs{z}} & \textrm{else}
  \end{cases}
\end{equation}
to the complex plane.
This complex sign function is the difference between the gpc- and the pc-generator.
The flow terminates when the generator vanishes, which is the case when only qp-conserving terms are left in the Lindbladian.

The flowing Lindbladian $\Lindbladian(\ell)$ depends on the flow parameter $\ell$ as well, which allows us to transform the original Lindbladian $\Lindbladian = \Lindbladian(0)$ into the effective Lindbladian $\Lindbladian_\textrm{eff} = \Lindbladian(\infty)$.
It is possible to circumvent calculating $S(\ell)$ explicitly by using the flow equation
\begin{equation}
  \partial_\ell \Lindbladian(\ell) = [\eta(\ell), \Lindbladian(\ell)],\label{eq:flow-L}
\end{equation}
which is equivalent to Eq.~\eqref{eq:flow-S}.
The gpc-generator leads to a block-diagonal effective Lindbladian that conserves the number of qps.
The flow equation is usually not exactly solvable, because it can consist of infinitely many coupled differential equations for the prefactors of operators in $\Lindbladian(\ell)$.
In the next two sections, we present two different perturbative truncation schemes to obtain finite closed sets of equations.
Both of them lead to effective Lindbladians that are valid in the thermodynamic limit.

\subsection{\texorpdfstring{\pcst}{pcst++}}
\label{sec:pcst}

The \pcst is introduced in~\cite{LenkeSchellenbergerSchmidt2023} using a general notation.
We repeat the formalism here, but with a notation adapted to the DTFIM.

The \pcst is applicable when two requirements are fulfilled.
Firstly, the unperturbed Lindbladian $Q$ can be written as the sum of two commuting counting operators
\begin{equation}
  Q^{\pm} = \left(\pm 2 h - \frac{\ii \Gamma}{2}\right) \sum_j \crealr{j} \annilr{j},
\end{equation}
which have an equidistant spectrum.
Secondly, the perturbation $V$ can be written as the sum of operators $T_{(m^+, m^-)}$ that fulfill
\begin{equation}
  [Q^{\pm}, T_{(m^+, m^-)}] = m^\pm T_{(m^+, m^-)}.
\end{equation}
The perturbation operators are listed in Tab.~\ref{tab:perturbation}.
These operators are the only building blocks that can make up the flowing Lindbladian $\Lindbladian(\ell)$.
We therefore make the perturbative ansatz
\begin{equation}
  \Lindbladian(\ell) = Q + \sum_{n = 1}^\infty (- J)^n \sum_{\abs{\mathbf{m}} = n} F(\ell; \mathbf{m}) T_\mathbf{m}
\end{equation}
with $T_\mathbf{m} = T_{(m^+_1, m^-_1)} \cdots T_{(m^+_n, m^-_n)}$ and $\mathbf{m} = [(m^+_1, m^-_1), \dots, (m^+_n, m^-_n)]$.
The specific form of the ansatz is determined solely by the complex-valued coefficient functions $F$.
The expression $\abs{\mathbf{m}} = n$ means that the sequence $\mathbf{m}$ consists of $n$ pairs $(m^+_i, m^-_i)$.
Using this ansatz, the generator~\eqref{eq:generator} reads
\begin{align}
  \eta(\ell) &= \sum_{n = 1}^\infty (- J)^n \sum_{\abs{\mathbf{m}} = n} \sgn(M(\mathbf{m})) F(\ell; \mathbf{m}) T_\mathbf{m},\\
  M(\mathbf{m}) &= \left(2 h - \frac{\ii \Gamma}{2}\right) \sum_{i = 1}^n m^+_i + \left(- 2 h - \frac{\ii \Gamma}{2}\right) \sum_{i = 1}^n m^-_i.
\end{align}
The two prefactors are independent, so $M(\mathbf{m})$ is vanishing exactly when no qps are created or annihilated, i.e., $\sum_i m^+_i = \sum_i m^-_i = 0$.
Here we see that Lindbladians conserving both qp types individually are fixed points of the flow.
And this makes it easier to determine which terms occur in the generator; especially for our computer implementation.
Inserting the ansatz for $\Lindbladian(\ell)$ and the expression for $\eta(\ell)$ into the flow equation~\eqref{eq:flow-L} leads to the differential equations
\begin{equation}
  \partial_\ell f(\ell; \mathbf{m}) = \sum_{(\mathbf{m}_a, \mathbf{m}_b) = \mathbf{m}} e^{(\abs{M(\mathbf{m})} - \abs{M(\mathbf{m}_a)} - \abs{M(\mathbf{m}_b)}) \ell} \left[\sgn(M(\mathbf{m}_a)) - \sgn(M(\mathbf{m}_b))\right] f(\ell; \mathbf{m}_a) f(\ell; \mathbf{m}_b)
\end{equation}
for the modified coefficient functions $f(\ell; \mathbf{m}) = e^{\abs{M(\mathbf{m})} \ell} F(\ell; \mathbf{m})$.
The expression $(\mathbf{m}_a, \mathbf{m}_b) = \mathbf{m}$ references all possible splittings of the sequence $\mathbf{m}$ into two smaller non-empty subsequences $\mathbf{m}_a$ and $\mathbf{m}_b$.
This is a set of recursive differential equations, that can be solved order by order with initial conditions $f(0; \mathbf{m}) = \delta_{1, \abs{\mathbf{m}}}$.
The resulting effective Lindbladian is obtained by taking the limit $\ell \to \infty$.
It is guaranteed that this limit exists, because of the proofs by induction in the appendix of~\cite{LenkeSchellenbergerSchmidt2023}.
It is of the form
\begin{equation}
  \Lindbladian_\textrm{eff} = Q + \sum_{n = 1}^\infty (- J)^n \sum_{\substack{\abs{\mathbf{m}} = n \\ M(\mathbf{m}) = 0}} F(\infty; \mathbf{m}) T_\mathbf{m}, \label{eq::Leff}
\end{equation}
because all terms with $M(\mathbf{m}) \neq 0$ decay during the flow.
The fact that only terms with $M(\mathbf{m}) = 0$ remain shows that the effective Lindbladian is block-diagonal and qp conserving.

It is easy to calculate the first two orders by hand, but for higher orders a computer implementation is necessary.
The program we used for calculating the prefactors $F(\infty; \mathbf{m})$ is available online~\cite{LenkeSchellenbergerSchmidt2023-software}.

The effective Lindbladian \eqref{eq::Leff} is not yet normal-ordered, so it is not immediately possible to investigate individual qp-sectors.
Normal ordering using the commutation relations between the perturbation operators is cumbersome, because they do not have simple commutation relations.
Instead, we exploit the linked-cluster property and calculate matrix elements on sufficiently large finite clusters.
Because of the qp conservation we can obtain the result for the effective Lindbladian in the n-qp sector by calculating all matrix elements in that sector and then subtracting all processes that already occurred for smaller qp numbers.
This is done by subtracting the same matrix element on the same finite cluster with one qp less if that qp did not move and by additionally subtracting the 0-qp prefactor from all matrix elements.

Let us exemplify the cluster calculations for the one-dimensional DTFIM in order 7 perturbation theory.
To this end we use a pictorial representation for the states given as finite segments of the ladder illustrated in Fig.~\ref{fig:DTFIM-ladder}, where $\cdot$ represents empty sites and $\times$ represents sites occupied by a qp.
In the 0-qp sector, one needs to calculate the relevant matrix element on a single periodic cluster
\begin{equation}
  \Bra{\begin{smallmatrix} \cdot & \cdot & \cdot & \cdot & \cdot & \cdot & \cdot & \cdot \\ \cdot & \cdot & \cdot & \cdot & \cdot & \cdot & \cdot & \cdot \end{smallmatrix}} \Lindbladian_\textrm{eff} \Ket{\begin{smallmatrix} \cdot & \cdot & \cdot & \cdot & \cdot & \cdot & \cdot & \cdot \\ \cdot & \cdot & \cdot & \cdot & \cdot & \cdot & \cdot & \cdot \end{smallmatrix}}_{\rm p}.
\end{equation}
However, as outlined below, this matrix element vanishes to any order perturbation theory.
All calculations in sectors with a finite number of qps can be done on open clusters. 
For the 1-qp sector there are many non-vanishing operator sequences left in a given order.
As an example, the calculation needed to calculate the prefactor of $\sum_j \creal{j} \annil{j + 3}$ is
\begin{equation}
  \Bra{\begin{smallmatrix} \cdot & \cdot & \times & \cdot & \cdot & \cdot & \cdot & \cdot \\ \cdot & \cdot & \cdot & \cdot & \cdot & \cdot & \cdot & \cdot \end{smallmatrix}} \Lindbladian_\textrm{eff} \Ket{\begin{smallmatrix} \cdot & \cdot & \cdot & \cdot & \cdot & \times & \cdot & \cdot \\ \cdot & \cdot & \cdot & \cdot & \cdot & \cdot & \cdot & \cdot \end{smallmatrix}} - \Bra{\begin{smallmatrix} \cdot & \cdot & \cdot & \cdot & \cdot & \cdot & \cdot & \cdot \\ \cdot & \cdot & \cdot & \cdot & \cdot & \cdot & \cdot & \cdot \end{smallmatrix}} \Lindbladian_\textrm{eff} \Ket{\begin{smallmatrix} \cdot & \cdot & \cdot & \cdot & \cdot & \cdot & \cdot & \cdot \\ \cdot & \cdot & \cdot & \cdot & \cdot & \cdot & \cdot & \cdot \end{smallmatrix}}.
\end{equation}
An illustrative example for the 2-qp sector is the operator $\sum_j \creal{j} \crear{j - 1} \annil{j + 3} \annir{j - 1}$ whose prefactor is calculated as
\begin{equation}
  \Bra{\begin{smallmatrix} \cdot & \cdot & \cdot & \times & \cdot & \cdot & \cdot & \cdot & \cdot \\ \cdot & \cdot & \times & \cdot & \cdot & \cdot & \cdot & \cdot & \cdot \end{smallmatrix}} \Lindbladian_\textrm{eff} \Ket{\begin{smallmatrix} \cdot & \cdot & \cdot & \cdot & \cdot & \cdot & \times & \cdot & \cdot \\ \cdot & \cdot & \times & \cdot & \cdot & \cdot & \cdot & \cdot & \cdot \end{smallmatrix}} - \Bra{\begin{smallmatrix} \cdot & \cdot & \cdot & \times & \cdot & \cdot & \cdot & \cdot & \cdot \\ \cdot & \cdot & \cdot & \cdot & \cdot & \cdot & \cdot & \cdot & \cdot \end{smallmatrix}} \Lindbladian_\textrm{eff} \Ket{\begin{smallmatrix} \cdot & \cdot & \cdot & \cdot & \cdot & \cdot & \times & \cdot & \cdot \\ \cdot & \cdot & \cdot & \cdot & \cdot & \cdot & \cdot & \cdot & \cdot \end{smallmatrix}}.
\end{equation}
The sizes of these clusters are chosen just large enough to ensure that the qps have enough space to fluctuate left and right and still end up at the correct spot within seven orders.

\subsection{deepCST}
\label{sec:deepcst}

For the deepCST, like for deepCUT~\cite{KrullDrescherUhrig2012}, we do not use products of the operators $T_{(m^+, m^-)}$ as a basis, but normal-ordered products of the creation and annihilation operators, here labeled $B_x$.
Our ansatz for the flowing Lindbladian is
\begin{equation}
  \Lindbladian(\ell) = \sum_x G(\ell; x) B_x,
\end{equation}
which is a non-perturbative ansatz with general complex-valued coefficient functions $G$.
The size of this basis can be reduced drastically by using the symmetries of the Lindbladian that are listed in Sec.~\ref{sec:model-symmetries}.
The only symmetry that is strictly necessary for epCST and deepCST is the translational symmetry.
The generator~\eqref{eq:generator} expressed in the same basis reads
\begin{align}
  \eta(\ell) &= \sum_x \sgn(M(B_x)) G(\ell; x) B_x,\\
  M(B_x) &= \left(2 h - \frac{\ii \Gamma}{2}\right) n_x^+ + \left(- 2 h - \frac{\ii \Gamma}{2}\right) n_x^-,
\end{align}
where $n_x^\pm$ is the total amount of created qps of type $\pm$.
Here we see that Lindbladians conserving both qp types individually are fixed points of the flow.
Inserting the ansatz for $\Lindbladian(\ell)$ and the expression for $\eta(\ell)$ into the flow equation~\eqref{eq:flow-L} leads to the differential equations
\begin{equation}
  \partial_\ell G(\ell; x) = \sum_{y, z} D_{x y z} G(\ell; y) G(\ell; z), \label{eq:deepCST-flow}
\end{equation}
where the structure factors $D_{x y z}$ are determined by expanding the commutator
\begin{equation}
  [B_y, B_z] = \sum_x D_{x y z} B_x
\end{equation}
in the basis of the normal-ordered operators.

So far, this is an exact reformulation of the flow equation~\eqref{eq:flow-L} and thus not generally solvable.
The truncation scheme is based on a perturbative ansatz for the coefficient functions $G(\ell; x) = \sum_{n = 1}^\infty (- J)^n g^{(n)}(\ell; x)$.
Inserting this ansatz into Eq.~\eqref{eq:deepCST-flow} gives us the epCST flow equation
\begin{equation}
  \partial_\ell g^{(n)}(\ell; x) = \sum_{y, z} \sum_{p + q = n} D_{x y z} g^{(p)}(\ell; y) g^{(q)}(\ell; z).
\end{equation}
This set of recursive differential equations can be solved numerically order by order and yields the perturbative epCST expansion of the effective Lindbladian.
Exploiting the translational symmetry ensures that the basis can be reduced to a finite subset containing only the strictly necessary operators.
The basis can be reduced even further when targeting only specific quantities up to a given order $n$.
If we assign a minimal order $O_\textrm{min}(B_x)$ to every operator $B_x$, we can also assign a maximal order
by setting $O_\textrm{max}(B_x) = n$ for the targeted operators and recursively increasing
\begin{equation}
  O_\textrm{max}(B_x) = \max_{\{y, z | D_{x y z} \neq 0\}} \left( O_\textrm{max}(B_y) - O_\textrm{min}(B_z) \right)
\end{equation}
for all other operators until converged.
The reduced basis is then made up of all operators $B_x$ with $O_\textrm{min}(B_x) \leq O_\textrm{max}(B_x)$ and all structure factors $D_{x y z}$ with $O_\textrm{max}(B_x) < O_\textrm{min}(B_y) + O_\textrm{min}(B_z)$ are set to zero.
These reduction steps can only be done after the whole epCST flow equation with all structure factors have been set up.
It is however possible to reduce the calculation effort by using upper bounds for the maximal orders during the setup and omitting some terms before even calculating it.
The upper bound we use for the DTFIM is
\begin{equation}
  O_\textrm{max}(B_x) \leq n - \left\lceil \frac{\max(0, c_x^+ - 2)}{2} + \frac{\max(0, a_x^+ - 2)}{2} + \frac{\max(0, c_x^- - 2)}{2} + \frac{\max(0, a_x^- - 2)}{2} \right\rceil,
\end{equation}
where $c_x^\pm$ and $a_x^\pm$ are the amount of creation and annihilation operators of type $\pm$ in $B_x$.

The truncation for the deepCST formalism is defined by using the same reduced basis of normal-ordered operators $B_x$ and the same structure factors $D_{x y z}$ as for the epCST calculation and then solving the deepCST flow equation~\eqref{eq:deepCST-flow} numerically.
In contrast to the epCST calculation, all coefficient functions $G$ influence each other to infinite order.
We introduce the residual-off-diagonality (ROD)
\begin{equation}
  \textrm{ROD}(\ell) = \sqrt{\sum_{M(B_x) \neq 0} \abs{G(\ell; x)}^2},
\end{equation}
as a measure for the amount of non-diagonal terms in the flowing Lindbladian.
The flow is considered to have converged when $\textrm{ROD}(\ell) < 10^{-6}$ and diverged when $\textrm{ROD}(\ell) > 10^6$.
If converged, the result of the flow is a non-perturbative expansion of the effective Lindbladian that contains contributions from higher orders than the truncation order $n$.
Krull et al.~\cite{KrullDrescherUhrig2012} emphasize that reducing the basis before solving the deepCST flow equation is important for numerical stability and leads to a more accurate result.

It is possible to solve the flow for truncation orders up to $n = 2$ by hand, but for higher orders a computer implementation is necessary.
The effective Lindbladian is valid in the thermodynamic limit and normal-ordered.

\subsection{Structure of the effective Lindbladian}
\label{sec:blocks}

The effective Lindbladians obtained by both methods are very similar.
The deepCST Lindbladian consists only of normal-ordered operators and the \pcst Lindbladian can be normal ordered as explained in Sec.~\ref{sec:pcst}.
When expanding the prefactors of the deepCST operators up to a certain perturbation order, they are equal to the prefactors of the normal-ordered \pcst operators calculated up to the same order.
Both Lindbladians conserve the number of qps and can thus be decomposed into different parts
\begin{equation}
  \Lindbladian_\textrm{eff} = \sum_{b_+, b_- = 0}^\infty \Lindbladian_{b_+, b_-},
\end{equation}
where every operator in $\Lindbladian_{b_+, b_-}$ creates and annihilates exactly $b_\pm$ qps of type $\pm$.

For the long-time evolution it suffices to consider four of those parts.
The steady state is the sector without any qps, so all the information is in the part $\Lindbladian_{0, 0}$.
We are searching for the Liouvillian gap, so we need to calculate everything needed to construct the density matrix that is not the steady state but has the longest lifetime.
This density matrix needs one physical spin flip and is part of the sector with two qps of different types.
This means that the information is distributed into the four parts $\Lindbladian_{0, 0}$, $\Lindbladian_{1, 0}$, $\Lindbladian_{0, 1}$ and $\Lindbladian_{1, 1}$.
We are free to additionally consider coherences that are part of the two sectors with one qp and thus need $\Lindbladian_{0, 0}$, $\Lindbladian_{1, 0}$ and $\Lindbladian_{0, 1}$.
If we find that the Liouvillian gap closes in those two sectors then it also closes in the sector with two qps of different types.
The inverse is not guaranteed.
Eigenvectors from those sectors can be combined to form a valid density matrix that has trace one and is positive semi-definite.
For simplicity, we use the notation
\begin{align}
  \Lindbladian_{0, 0} = \Lindbladian_\textrm{0qp}, && \Lindbladian_{1, 0} = \Lindbladian_\textrm{1qp}^{(+)}, && \Lindbladian_{0, 1} = \Lindbladian_\textrm{1qp}^{(-)}, && \Lindbladian_{1, 1} = \Lindbladian_\textrm{2qp}
\end{align}
and call the operators from $\Lindbladian_\textrm{0qp}$ 0-qp operators, those from $\Lindbladian_\textrm{1qp}^{(\pm)}$ 1-qp operators and those from $\Lindbladian_\textrm{2qp}$ 2-qp operators.
Operators that create and annihilate exactly two qps of one type and no qps of the other are not relevant for us here, even though they could be considered 2-qp operators as well.

The identity operator is the only 0-qp operator and thus $\Lindbladian_\textrm{0qp}$ is just a number.
The 0-qp sector contains only the state without any qps.

The part containing the 1-qp operators is of the form
\begin{equation}
  \Lindbladian_\textrm{1qp}^{(\pm)} = f_0^\pm \sum_j \crealr{j} \annilr{j} + \sum_{\delta > 0} f_\delta^\pm \sum_j \left( \crealr{j} \annilr{j + \delta} + \crealr{j + \delta} \annilr{j} \right).
\end{equation}
Due to the translational symmetry, the prefactors $f_\delta$ do not depend on the site $j$, and because of the inversion symmetry, we have $f_{- \delta} = f_\delta$.
As a consequence of the particle swapping symmetry, the prefactors of the different qp types are related to each other by negation and complex conjugation
\begin{equation}
  f_\delta^\mp = - \left( f_\delta^\pm \right)^*.
\end{equation}
To further decompose this part into different sectors for different momenta, we perform a Fourier transformation
\begin{align}
  \crealr{k} &= \frac{1}{\sqrt{N}} \sum_j e^{\ii k j} \crealr{j}, & \annilr{k} &= \frac{1}{\sqrt{N}} \sum_j e^{- \ii k j} \annilr{j},
\end{align}
where $N$ is the number of spins in the chain in Fig.~\ref{fig:DTFIM-chain}, or, equivalently, dimers in the purified ladder in Fig.~\ref{fig:DTFIM-ladder}.
We obtain a diagonalized effective Lindbladian of the form
\begin{equation}
  \Lindbladian^{(\pm)}_\textrm{1qp} = \sum_k \omega^\pm(k) \crealr{k} \annilr{k}
\end{equation}
with
\begin{equation}
  \omega^\pm(k) = \sum_\delta \cos(k \delta) f_\delta^\pm.
\end{equation}
Because of the particle swapping symmetry, the eigenvalues of the two sectors are related via $\omega^\mp(k) = - \left( \omega^\pm(k) \right)^*$.
The resulting effective Lindbladian contains only terms where a qp moves at most $n$ sites for an $n$th order expansion, which is equivalent to saying $\abs{\delta} \leq n$.
To get the actual effective eigenvalue of a state with one qp, we add the 0-qp and 1-qp parts.
For the DTFIM the 0-qp contribution vanishes, so this is not necessary.

\begin{figure}[bt]
  \centering
  \includegraphics[page = 2, clip, trim = 20.7cm 7.3cm 17.8cm 12.5cm, width=0.8\textwidth]{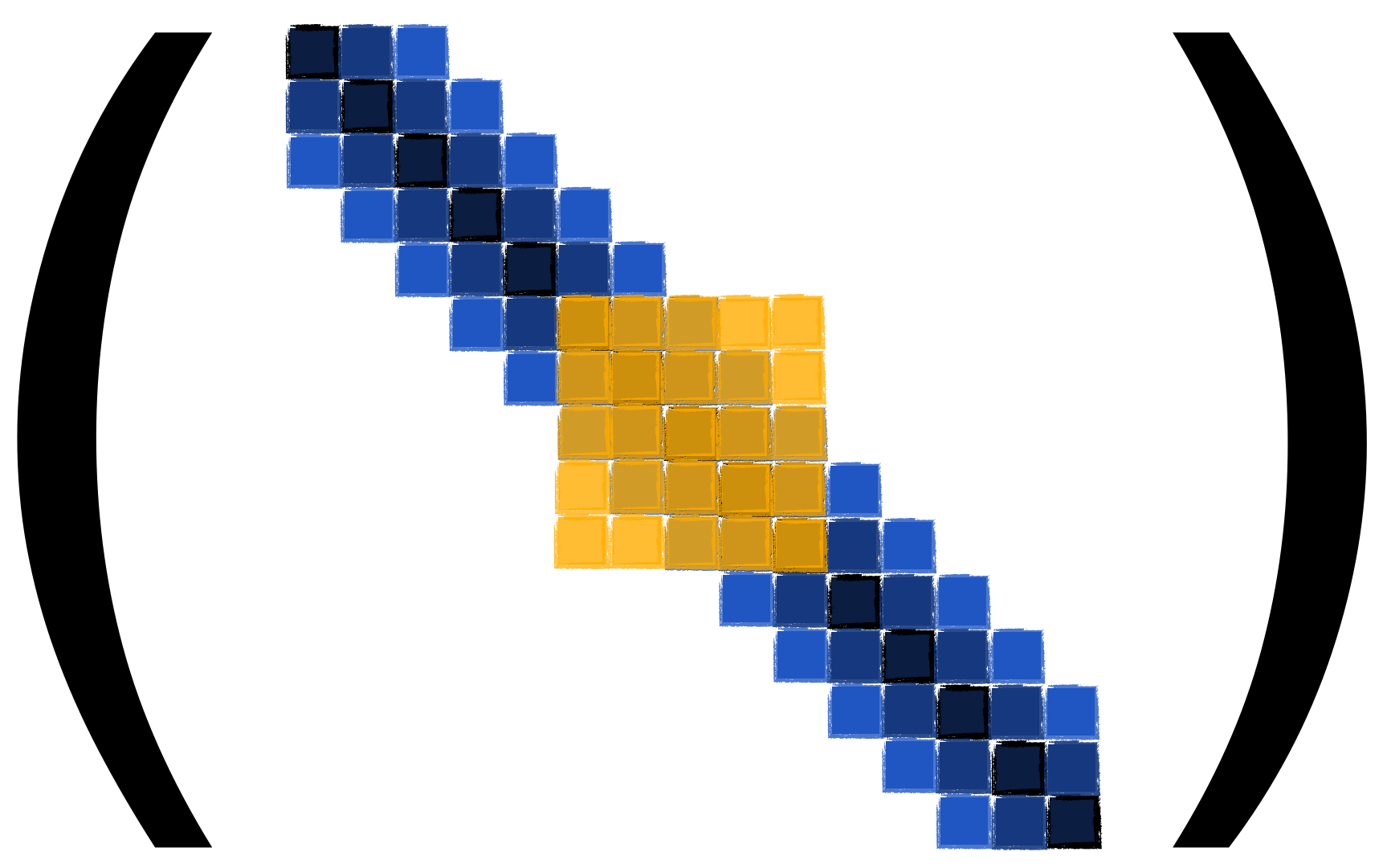}
  \caption{
    Pictorial representation of the matrix in the 2-qp sector after truncating the basis to \eqref{eq:2qp-basis} with $D = 7$ and $n = 3$.
    The blue diagonals contain the 1-qp operators and the yellow square contains the 2-qp operators.
  }
  \label{fig:2qp-matrix}
\end{figure}

The part containing the relevant 2-qp operators is of the form
\begin{equation}
  \begin{split}
  \Lindbladian_\textrm{2qp} &= \sum_{a, \gamma, \delta} g^a_{\gamma, \delta} \sum_j \creal{j + \gamma} \crear{j + a + \delta} \annil{j} \annir{j + a}.
  \end{split}
\end{equation}
Because of the translational symmetry, the prefactors $g^a_{\gamma, \delta}$ do not depend on the site $j$, but on the difference $a$ between the two qp types only.
Our convention for $a$ is the difference between the indices of the two annihilation operators.
The inversion symmetry gives us $g^a_{\gamma, \delta} = g^{- a}_{- \gamma, - \delta}$ and the particle swapping symmetry gives us
\begin{equation}
  g^a_{\gamma, \delta} = - \left( g^{- a}_{\delta, \gamma} \right)^*.
\end{equation}
To decompose this Lindbladian into different sectors for different momenta, we perform a Fourier transformation
\begin{align}
  \crea{K, d} &= \frac{1}{\sqrt{N}} \sum_j e^{\ii K (2 j + d) / 2} \creal{j} \crear{j + d}, & \anni{K, d} &= \frac{1}{\sqrt{N}} \sum_j e^{- \ii K (2 j + d) / 2} \annil{j} \annir{j + d}
\end{align}
to use composite particles with center of mass momentum $K$ and relative distance $d$.
The states with two qps are denoted by $\ket{K, d} = \crea{K, d} \ket{0}$, where $\ket{0}$ is the unperturbed stationary state with no qps.
We obtain an effective Lindbladian of the form
\begin{equation}
  \begin{split}
  \Lindbladian_\textrm{2qp} &= \sum_K \sum_{a, \gamma, \delta} g^a_{\gamma, \delta} e^{- \ii K (\gamma + \delta) / 2} \crea{K, a - \gamma + \delta} \anni{K, a},
  \end{split}
\end{equation}
that is not yet diagonalized, but decomposed into different sectors for different center of mass momenta.
The only non-vanishing operators are those where the two qps either hop past each other
\begin{equation}
  \{0, \dots, \gamma\} \cap \{a, \dots, a + \delta\} \neq \emptyset
\end{equation}
or where they are close enough together that they can feel each other via fluctuations
\begin{equation}
  2 \min(\abs{\max(0, \gamma) - \min(a, a + \delta)}, \abs{\max(a, a + \delta) - \min(0, \gamma)}) \leq n - \abs{\gamma} - \abs{\delta},
\end{equation}
where $n$ is the order of the truncation.
To get the actual effective eigenvalue of a state with two qps, we add the 0-qp, 1-qp and 2-qp parts.
In our case the 0-qp contribution vanishes, so this is not necessary.
The action of the 1-qp terms on a state $\ket{K, d}$ is
\begin{align}
  \sum_j \crealr{j} \annilr{j} \ket{K, d} &= \ket{K, d},\\
  \sum_j \creal{j} \annil{j + \delta} \ket{K, d} &= e^{\ii K \delta / 2} \ket{K, d + \delta},\\
  \sum_j \crear{j} \annir{j + \delta} \ket{K, d} &= e^{\ii K \delta / 2} \ket{K, d - \delta},\\
  \sum_j \creal{j + \delta} \annil{j} \ket{K, d} &= e^{- \ii K \delta / 2} \ket{K, d - \delta},\\
  \sum_j \crear{j + \delta} \annir{j} \ket{K, d} &= e^{- \ii K \delta / 2} \ket{K, d + \delta},
\end{align}
so they also leave the center of mass momentum invariant.
Then we fix the $K$ value, truncate the relative distance $d$ at some maximum value $D$ and write down the resulting matrix in the basis
\begin{equation}
  \{\ket{K, - D}, \ket{K, - D + 1}, \dots, \ket{K, - 1}, \ket{K, 0}, \ket{K, 1}, \dots, \ket{K, D - 1}, \ket{K, D}\}, \label{eq:2qp-basis}
\end{equation}
as visualized in Fig.~\ref{fig:2qp-matrix}.
The 1-qp operators act independently of $d$ and their prefactors are located in a band with width $2 n + 1$ around the diagonal.
The prefactors of the 2-qp operators are located in a $(2 n + 1) \times (2 n + 1)$ block centered around $\ket{K, 0}$.
The resulting finite matrix can be numerically diagonalized to obtain the eigenvalues for this fixed $K$ value.
Despite the truncation in the distance between the two qps, the results are still valid in the thermodynamic limit.

The form of all Lindbladian parts does not depend on the method used.
The difference between \pcst and deepCST is in the prefactors only.

\section{Results}
\label{sec:results}
\begin{figure}[btp]
\centering
  \resizebox{0.9\textwidth}{!}{
  \input{Plots/1qp_real_dispersion.tex}}
  \caption{
    Real part of the eigenvalues $\pm \omega^\pm(k)$ in the 1-qp sector for the parameters $h = \Gamma = 1$ for $k \in\{ 0, \pi\}$ as a function of $J$.
    The lines represent the \pcst expansion up to different orders.
    The colored solid lines represent $k = 0$ and the colored dotted lines $k = \pi$.
    The respective thin black lines represent the highest diagonal Padé approximants of the highest order \pcst series.
    The scatter plots represent the deepCST expansion up to different orders.
    The bigger black markers represent $k = 0$ and the smaller gray markers $k = \pi$.
    Orders lower than $4$ are not plotted.
    \label{fig:1qp_real_dispersion}
  }
\vspace*{+5mm}
  \resizebox{0.9\textwidth}{!}{
  \input{Plots/1qp_real_dispersion_k.tex}}
  \caption{
    Real part of the eigenvalues $\pm \omega^\pm(k)$ in the 1-qp sector for the parameters $h = \Gamma = 1$ for different values of $J$ as a function of $k$.
    The thick (thin) solid lines represent the \pcst expansion up to \nth{7} (\nth{6}) order.
    At $J = 0.3$ those two orders start to differ.
    The thick (thin) dotted lines represent the deepCST expansion up to \nth{6} (\nth{5}) order.
    At $J = 0.3$ those two orders start to differ.
    \label{fig:1qp_real_dispersion_k}
  }
\end{figure}
We calculated the effective Lindbladian using the three methods \pcst, epCST and deepCST.
The epCST results are identical by definition to the \pcst results up to the calculated order.

\subsection{0 qp: Stationary state}
\label{sec:0-qp}
As already mentioned in~\cite{LenkeSchellenbergerSchmidt2023}, the left eigenstate without any qps is also a left eigenstate of all perturbation operators with eigenvalue $0$.
Because every operator sequence in the effective Lindbladian obtained via the \pcst calculation consists of these operators only, the state without any qps is also a left eigenstate of the effective Lindbladian with eigenvalue $0$.
This proves that the corresponding right eigenstate is a stationary state independent of the chosen parameters.
The same must hold for the deepCST calculation. 
To summarize, we have that
\begin{equation}
  \Lindbladian_\textrm{0qp} = 0
\end{equation}
up to arbitrary order.
Thus, the state without any qps is a stationary state for all $J$ in the convergence radius of the perturbative expansion, which we have confirmed numerically up to the calculated orders.

\subsection{1 qp: Coherences}
\label{sec:1-qp}

The results for $\omega^\pm(k)$ obtained by the \pcst up to order $n = 7$ and by deepCST up to order $n = 6$ are displayed in Figs.~\ref{fig:1qp_real_dispersion} and \ref{fig:1qp_imag_dispersion} as functions of $J$ for selected values of $k$ and in Figs.~\ref{fig:1qp_real_dispersion_k} and \ref{fig:1qp_imag_dispersion_k} as functions of $k$ for selected values of $J$.
The other parameters are $h = \Gamma = 1$.
\begin{figure}[btp]
  \centering
  \resizebox{0.9\textwidth}{!}{\input{Plots/1qp_imag_dispersion.tex}}
  \caption{Imaginary part of the eigenvalues $\pm \omega^\pm(k)$ in the 1-qp sector for the parameters $h = \Gamma = 1$ for $k \in\{ 0, \pi/2\}$ as a function of $J$.
    The lines represent the \pcst expansion up to different orders.
    The colored solid lines represent $k = 0$ and the colored dotted lines $k = \pi/2$.
    The scatter plots represent the deepCST expansion up to different orders.
    The bigger black markers represent $k = 0$ and the smaller gray markers $k = \pi/2$.
    Orders lower than $4$ are not plotted.
    \label{fig:1qp_imag_dispersion}
  }
\vspace*{+5mm}
  \resizebox{0.9\textwidth}{!}{
  \input{Plots/1qp_imag_dispersion_k.tex}}
  \caption{
    Imaginary part of the eigenvalues $\pm \omega^\pm(k)$ in the 1-qp sector for the parameters $h = \Gamma = 1$ for different values of $J$ as a function of $k$.
    The thick (thin) solid lines represent the \pcst expansion up to \nth{7} (\nth{6}) order.
    At $J = 0.18$ the leading order of the \pcst expansion starts to invert its maxima and minima with respect to smaller $J$.
    The thick (thin) dotted lines represent the deepCST expansion up to \nth{6} (\nth{5}) order.
    At $J = 0.4$ the leading order of the deepCST expansion starts to invert its maxima and minima with respect to smaller $J$ (not shown).
    \label{fig:1qp_imag_dispersion_k}
  }
\end{figure}

The real part of the \pcst series is converged up to $J \approx 0.3$ as indicated by a visible difference between the two leading orders in Fig.~\ref{fig:1qp_real_dispersion_k}.
The imaginary part of the \pcst series is converged up to $J \approx 0.16$ as indicated by the inversion of its maxima and minima with respect to smaller $J$ in Fig.~\ref{fig:1qp_imag_dispersion_k}.
The highest diagonal Padé approximants of the real part of the \pcst series are also displayed in Fig.~\ref{fig:1qp_real_dispersion}.
The lower diagonal approximants are very similar for the whole plotted region, indicating that the convergence radius of the real part of the series can be greatly increased.
The Padé approximants for the imaginary part on the other hand differ visibly also for small $J$.
This is not displayed in Fig.~\ref{fig:1qp_imag_dispersion} for visual clarity.

In comparison, the deepCST results are converged for larger values of $J$ than the \pcst results and the different orders lie closer together also for larger $J$.
Inside their convergence radius, both methods lead to the same results.
Comparing the Padé approximants of the real part of the \pcst results with the deepCST results shows that they are very similar, indicating that both methods lead to the same results also outside the convergence radius of the \pcst.

\subsection{2 qp: One spin flip}
\label{sec:2-qp}

\begin{figure}[bt]
  \input{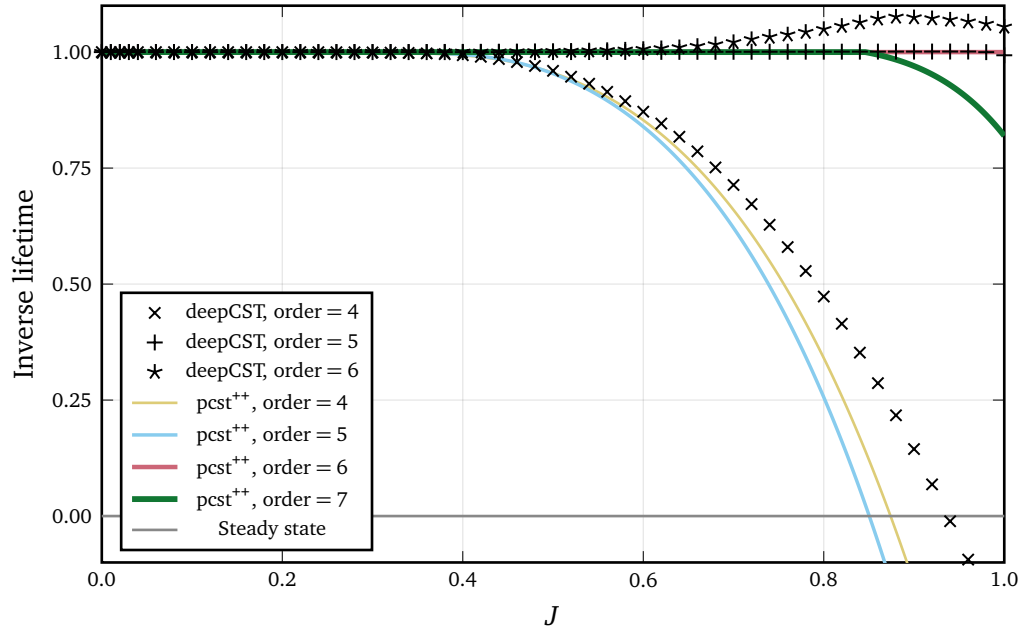}
  \caption{
    Inverse lifetime of the longest-living state in the 2-qp sector for the parameters $h = \Gamma = 1$ and a maximum relative distance of $D = 50$ as a function of $J$.
    The longest-living state belongs to the state with center of mass momentum $K = 0$ and its inverse lifetime is given by $- \im(\lambda)$, where $\lambda$ is its eigenvalue.
    The solid lines represent the \pcst expansion up to different orders.
    The scatter plots represent the deepCST expansion up to different orders.
    All lower orders are flat in both expansions.
    \label{fig:2qp_dispersion}
  }
\end{figure}

The eigenvalue with the largest imaginary part, i.e., with the longest lifetime, belongs to the state with center of mass momentum $K = 0$, as indicated by both methods.
The \pcst results up to order $n = 7$ and the deepCST results up order $n = 6$, both with a maximum relative distance of $D = 50$, are displayed in Fig.~\ref{fig:2qp_dispersion} as a function of $J$.
The parameters are $h = \Gamma = 1$.

Both expansions are converged and agree up to $J \approx 0.4$.
Choosing a bigger maximum relative distance than $D = 50$ would not change the results.
Within this convergence radius, the imaginary part of this eigenvalue stays relatively constant, which implies that the Liouvillian gap does not close and that there is no steady-state phase transition.
This is in accordance to the literature~\cite{JoshiNissenKeeling2013,FossFeigYoungAlbertGorshkovMaghrebi2017,AliKamarSeifMaghrebi2026}.
Outside the convergence radius, there are eigenvalues whose imaginary part approaches $0$, but these results are inconclusive within the achieved order of expansion.

\section{Conclusion}
\label{sec:conclusion}

We have discussed two general frameworks for deriving effective quasiparticle-conserving Lindbladian operators directly in the thermodynamic limit.
Our approaches are both extensions of continuous unitary transformations to open quantum systems and thereby establish a systematic route to block-diagonalize effective Lindbladians in a quasiparticle basis.
Consequently, one has transformed the open quantum many-body problem into an effective few-body problem corresponding to the long time evolution.
Both methods use the gpc-generator~\cite{SchmiedinghoffUhrig2022}.
The perturbative \pcst approach is an extension of perturbative continuous unitary transformations (pCUT) and combines quasiparticle-conserving transformations with linked-cluster expansions, enabling high-order series expansions directly for the infinite system.
The non-perturbative deepCST approach extends directly evaluated enhanced perturbative continuous unitary transformations (deepCUT) to Lindbladian operators through the same generator and a perturbative truncation scheme, also providing effective, quasiparticle-conserving Lindbladian operators without finite-size limitations.
Conceptionally, we aim at isolating the steady-state (absence of quasiparticles in the effective description) and long live-time excited states in the low-quasiparticle parts of the effective Lindbladian, so that we have access to the physical properties of steady-state phases and their critical breakdown, e.g., by exploring the Liouvillian gap.

The \pcst method~\cite{LenkeSchellenbergerSchmidt2023} is a generalization of pCUT and thus a perturbative approach~\cite{KnetterUhrig2000,KnetterSchmidtUhrig2003a}.
One obtains model-independently an effective, quasiparticle-conserving Lindbladian as a high-order perturbative expansion, which is, however, not normal-ordered.
The normal-ordering is model-dependent and most efficiently done via a linked-cluster expansion on topologically distinct graphs.
As a consequence, one can reach similar maximal perturbative orders in any spatial dimension with this approach, which is a very attractive property.
As for closed systems, extrapolation tools like Padé or DLog Padé approximants can enhance the convergence radius of the series and allow in principle to access critical regimes.

In contrast, the deepCST method, which we introduced in this work, is a generalization of deepCUT~\cite{KrullDrescherUhrig2012}.
The deepCST operates directly on the operator level in the thermodynamic limit.
The resulting effective Lindbladian is normal-ordered and is derived non-perturbatively by numerically solving the flow equation, which is truncated by a perturbative truncation scheme.
As a consequence, \pcst and deepCST results must be identical up to the common maximal perturbative order of the calculation, i.e., for small values of the perturbation.
At the same time, the deepCST is expected to converge in a larger coupling domain due to its non-perturbative character.

We have applied both methods to the dissipative transverse field Ising model in one dimension, with the Ising interaction as the perturbation.
As a preparatory step the Lindbladian of the model has been purified by splitting each spin into a dimer of two sites -- one for the bra and one for the ket of its density matrix.
In this representation one obtains a two-leg ladder geometry so that one physical spin flip corresponds to two quasiparticles on the same rung of the ladder.
The \pcst calculations have been performed up to order 7 in perturbation theory in all sectors up to two quasiparticles while we reached order 6 for the same quasiparticle sectors in the deepCST approach.
The state without any quasiparticles is the stationary steady state, which is shown to hold true in any order of perturbation theory.
To extract physical properties of quasiparticles, one has to diagonalize the quasiparticle sectors of interest.
Here we calculated the eigenvalues of every 1-quasiparticle state in momentum representation exploiting the translational symmetry.
In addition, we determined the eigenvalues of the 2-quasiparticle sector using exact diagonalization in a basis with fixed center of mass momentum $K$ and truncated relative distance between the two quasiparticles.
The 2-quasiparticle state whose eigenvalue has the largest imaginary part corresponds to the Liouvillian gap and has center of mass momentum $K = 0$.
All our findings confirm the absence of a phase transition in the dissipative transverse field Ising model in one dimension~\cite{JoshiNissenKeeling2013,FossFeigYoungAlbertGorshkovMaghrebi2017,AliKamarSeifMaghrebi2026}.

In the future it would be interesting to extend the investigation of open quantum many-body systems to models where one expects different phases and phase transitions in the steady-state phase diagram. 
One obvious next step is to study the dissipative transverse field Ising model in higher dimension, e.g., on the two-dimensional square lattice where a phase transition has been detected numerically~\cite{MaghrebiGorshkov2016,OverbeckMaghrebiGorshkovWeimer2017,JinBiellaViyuelaCiutiFazioRossini2018,JinHeIeminiFerreiraWangChesiFazio2021}.  
Altogether, \pcst and deepCST considerably broaden the toolbox for studying dissipative quantum many-body systems in terms of their elementary quasiparticle excitations directly in the thermodynamic limit.

\section*{Acknowledgements}
We thank Andreas Schellenberger for fruitful discussions.

\paragraph{Author contributions}
LL: Conceptualization, Data curation, Formal analysis, Investigation, Methodology, Software, Visualization, Writing -- original draft, Writing -- review \& editing.
KPS: Conceptualization, Funding acquisition, Methodology, Resources, Supervision, Writing -- review \& editing.\footnote{Following the taxonomy \href{https://credit.niso.org}{CRediT} to categorize the contributions of the authors.}

\paragraph{Funding information}
LL and KPS acknowledge support by the Deutsche Forschungsgemeinschaft (DFG, German Research Foundation), Project-ID 429529648-TRR 306 QuCoLiMa (Quantum Cooperativity of Light and Matter) and by the Munich Quantum Valley, which is supported by the Bavarian state government with funds from the Hightech Agenda Bayern Plus.

\bibliography{bib.bib}

\end{document}